\documentclass[10pt,english]{article}
\usepackage{subfigure}
\usepackage{multirow}
\usepackage{graphicx}
\usepackage{float}
\usepackage[T1]{fontenc}
\usepackage[latin9]{inputenc}
\usepackage[margin=1.25in]{geometry}
\usepackage{float}
\usepackage{amsmath}
\usepackage{amsbsy}
\usepackage{setspace}
\usepackage{amssymb}
\usepackage{esint}
\usepackage{cite}
\usepackage{hyperref}
\newfloat{algorithm}{tbp}{loa}
\floatname{algorithm}{Algorithm}
\usepackage{algorithmic}

\newlength\myindent
\makeatletter

\floatstyle{ruled}
\newfloat{algorithm}{tbp}{loa}
\floatname{algorithm}{Algorithm}

\usepackage{babel}

\begin{document}

\title{Attack vs Benign Network Intrusion Traffic Classification}

\author{M. Andrecut}

\date{May 6, 2022}

\maketitle
{

\centering Calgary, Alberta, Canada

\centering mircea.andrecut@gmail.com

} 
\begin{abstract}
Intrusion detection systems (IDS) are used to monitor networks or systems for attack activity or policy violations. 
Such a system should be able to successfully identify anomalous deviations from normal traffic behavior. 
Here we discuss the machine learning approach to building an anomaly-based IDS using the CSE-CIC-IDS2018 dataset. 
Since the publication of this dataset a relatively large number of papers have been published, most of them presenting IDS architectures and results based 
on complex machine learning methods, like deep neural networks, gradient boosting classifiers, or hidden Markov models. Here we show that similar results can be obtained using a 
very simple nearest neighbor classification approach, avoiding the inherent complications of training such complex models. 
The advantages of the nearest neighbor algorithm are: (1) it is very simple to implement; (2) it is 
extremely robust; (3) it has no parameters, and therefore it cannot overfit the data. This result also shows that currently 
there is a trend of developing over-engineered solutions in the machine learning community. Such solutions are based on complex methods, like deep learning neural networks, 
without even considering baseline solutions corresponding to simple, but efficient methods.  
\smallskip 

Keywords: intrusion detection systems, machine learning, deep learning, nearest neighbor.

\end{abstract}

\section{Introduction}

Anomaly detection in network traffic is an important aspect of information security, and a frequently used method for detecting zero day attacks. 
Intrusion detection systems (IDS) are required to monitor networks or systems in order to detect anomalous patterns caused by network attacks, 
using machine learning algorithms \cite{key-1}-\cite{key-21}.

IDS are typically classified into two categories, according to their detection type: (1) signature-based; and (2) anomaly-based methods. 
A signature-based IDS is similar to anti-virus software that can detect malicious patterns known as signatures.
This type of IDS has high accuracy and low false-positive rate for known attacks, but they have no mechanism to detect novel attacks. 
The anomaly-based IDS uses machine learning techniques to detect anomalous traffic. This type of IDS usually has a higher false-positive rate, 
and the inconvenience of training a model, which in turn requires a good training data set \cite{key-1}-\cite{key-20}. 

Here we discuss the machine learning approach to building an anomaly-based IDS using the CSE-CIC-IDS2018 dataset \cite{key-21}. 
Since its publication, the CSE-CIC-IDS2018 dataset has been frequently used to train IDS based on machine learning methods,
and it has been adopted as a benchmark for anomaly-based IDS implementations \cite{key-1}-\cite{key-20}\footnote{There may be also other papers we are not aware of, 
we apologise for not mentioning them here.}. Interestingly, most of these implementations 
are based on complex solutions like deep learning neural networks, random forest and gradient boosting classifiers, or hidden Markov models. 
While all these implementations report very good results, deep learning methods stand out by reporting a stellar 0.99 accuracy. However, all these 
papers fall short in comparing their results to other methods. In contrast to the above papers, here we show that similar results can be obtained using a 
very simple nearest neighbor classification approach, avoiding the inherent complications of training deep learning neural networks or other complex classifiers. 
The advantages of the nearest neighbor algorithm are: (1) it is very simple to implement; (2) it is 
extremely robust; (3) it has no parameters, and therefore it cannot overfit the data. 

Our investigation also shows that currently there is a trend of developing over-engineered solutions in the machine learning community. Such solutions are based on complex methods, like deep learning neural networks, 
without even considering baseline solutions corresponding to simple, but efficient methods.

\section{The CSE-CIC-IDS2018 dataset}

The CSE-CIC-IDS2018 dataset was published by the Communications Security Establishment (CSE) and the Canadian Institute for Cybersecurity (CIC) \cite{key-21}. 
The CSE-CIC-IDS2018 dataset includes benign samples and samples corresponding to several different attack scenarios, including: Brute-force, Denial of Service (DoS), Web attacks, Botnet, and Infiltration.

The data is organized in seven CSV files, where each row is a sample, labeled as benign or with the name of the corresponding attack, and it consists on 80 traffic features extracted using CICFlowMeter \cite{key-22}. 
After downloading the data, we cleaned the data using the Python script provided by \cite{key-23}. That is, we dropped the samples with missing feature values, and removed the columns with no values. In addition, 
we also dropped the timestamp, since it doesn't play any role in the classification. Table 1 shows the statistics summary of the data after the cleaning process. 

\begin{table}[h!]
\begin{center}
\caption{Summary of the CSE-CIC-IDS2018 dataset.}
\bigskip
\begin{tabular}{ |c|c|c| } 
\hline
Data File & Traffic Type & Number of Samples \\ 
\hline \multirow{3}{8em}{02-14-2018.csv} & Benign & 663,808 \\ & FTP-BruteForce & 193,354 \\ & SSH-Bruteforce & 187,589 \\ 
\hline \multirow{3}{8em}{02-15-2018.csv} & Benign & 988,050 \\ & DoS-GoldenEye & 41,508 \\ & DoS-Slowloris & 1,099 \\
\hline \multirow{3}{8em}{02-16-2018.csv} & Benign & 446,772 \\ & DoS-SlowHTTPTest & 139,890 \\ & DoS-Hulk & 461,912 \\
\hline \multirow{3}{8em}{02-22-2018.csv} & Benign & 1,042,603 \\ & BruteForce-Web & 249 \\ & BruteForce-XSS & 79 \\
\hline \multirow{4}{8em}{02-23-2018.csv} & Benign & 1,042,301 \\ & BruteForce-Web & 362 \\ & BruteForce-XSS & 151 \\ & SQL-Injection & 53 \\
\hline \multirow{2}{8em}{03-01-2018.csv} & Benign & 235,778 \\ & Infiltration & 92,403 \\
\hline \multirow{2}{8em}{03-02-2018.csv} & Benign & 758,334 \\ & BotAttack & 286,191 \\
\hline \multirow{2}{8em}{Binary Class} & Benign & 5,177,646 \\ & Attack & 1,414,765 \\ \hline
\end{tabular}
\end{center}
\end{table}

\section{Data normalization}

Our intention is to use the 5-fold validation approach. So, for each file we split the data in 5 equal parts, we use 4 parts (80\%) for "training" and 1 part (20\%) for "testing". 
That is, for each data file we use 5 "training-testing" iterations, and  we take the average of the metrics (precision, recall, accuracy, and the F-measure). 
Let us assume that $D = \lbrace d_j \vert j=1,...,J \rbrace$ is the data corresponding to a data file, 
 $Y = \lbrace y_n \vert n=1,...,N \rbrace$ is the "training" set and $X = \lbrace x_m \vert m=1,...,M \rbrace$ is the "testing" set in such an iteration. 
We also assume that $C_D = \lbrace c_j \vert j=1,...,J \rbrace$ are the binary labels corresponding to $\lbrace benign, attack\rbrace \equiv \lbrace 0, 1\rbrace$. 
Consequently we have $C_Y = \lbrace c^Y_n \vert n=1,...,N \rbrace$ and $C_X = \lbrace c^X_m \vert m=1,...,M \rbrace$ the binary labels of the "training", and respectively 
"testing" data sets. 
Obviously for each of the 5-fold iterations we have: $J=N+M$, $N/J\simeq0.8$, $M/J\simeq0.2$, and $D=X\cup Y$, $C_D=C_X \cup C_Y$. 

As mentioned above, each data row is represented as a vector with $K$ columns, corresponding to the features extracted by the CICFlowMeter: 
\begin{equation}
d_{jk} = [d_{j1},...,d_{jK}], j=1,...,J.
\end{equation}
In order to use the "dot product similarity" as a metric for the nearest neighbor algorithm we have to normalize the data such that each $d_j$ vector becomes a unit 
vector in the $K$-dimensional feature space. The normalization consists of two steps. In the first step we standardize each column using the Z-score approach. 
That is for each column $k$ we calculate the mean $\mu_k$ and standard deviation $\sigma_k$ an we normalize the column as following:
\begin{equation}
d_{jk} \longleftarrow (d_{jk} - \mu_k)/\sigma_k,\quad j=1,...,J, k=1,...,K.
\end{equation}
The second step consists in normalizing the row vectors using the Euclidean norm:
\begin{equation}
d_{j} \longleftarrow d_j/\Vert d_j \Vert,\quad j=1,...,J.
\end{equation}

\section{The Nearest Neighbor algorithm}
After the normalization, all the data rows have the property $\Vert d_j \Vert=1$, and therefore the "similarity" $s(d_i,d_j)$ between two vectors $d_i$ and $d_j$ can be easily measured 
using the dot product:
\begin{equation}
s(d_i,d_j) = \langle d_i,d_j \rangle = \Vert d_i \Vert \Vert d_j \Vert \cos (d_i,d_j) = \cos (d_i,d_j) \in [-1,1].
\end{equation}
Thus, two vectors are "similar" if they are colinear, such that the cosine of their angle is $\cos (d_i,d_j)=1$, also they are dissimilar if they are orthogonal, that is $\cos (d_i,d_j)=0$. 
Obviously, for $\cos (d_i,d_j)=-1$ the vectors are pointing in opposing directions. 

In order to "classify" a "test" sample $x_m$ we simply calculate the matrix-vector product:
\begin{equation}
p_m = Y x_m = [p_{m1},...,p_{mN}].
\end{equation}
The result $p_m$ is a vector of length $N$ corresponding to the projection of $x_m$ onto all the vectors from matrix $Y$ with the rows corresponding to the "training" data set. 
Then we select the index of the element with the maximum value in the vector $p_m$:
\begin{equation}
n = \text{arg} \max_n [p_{m1},...,p_{mN}]. 
\end{equation}
After the index $n$ is determined, the classification of $x_m$ consists in assigning it the label $c^Y_n$ from the "training" data set. 
The result of the classification is:
\begin{itemize}
\item True positive (TP) if:
\begin{equation}
c^Y_n = c^X_m = 1,
\end{equation}
\item True negative (TN) if:
\begin{equation}
c^Y_n = c^X_m = 0,
\end{equation}
\item False positive (FP) if:
\begin{equation}
c^Y_n = 1 \: \text{and} \: c^X_m = 0,
\end{equation}
\item False negative (FN) if:
\begin{equation}
c^Y_n = 0 \: \text{and} \: c^X_m = 1.
\end{equation}
\end{itemize}

In order to characterize the performance of the nearest neighbor classifier we use the following metrics averaged over the 5-fold iterations:
\begin{itemize}
\item Precision: 
\begin{equation}
P = \frac{TP}{TP+FP}.
\end{equation}
\item Recall: 
\begin{equation}
R = \frac{TP}{TP+FN}.
\end{equation}
\item Accuracy: 
\begin{equation}
A = \frac{TP+TN}{TP+FP+TN+FN}.
\end{equation}
\item F-measure: 
\begin{equation}
F = 2\frac{PR}{P+R}.
\end{equation}
\end{itemize}

\section{Numerical Results}

The obtained results for the nearest neighbor algorithm are given in Table 2. 
\begin{table}[h!]
\begin{center}
\caption{Nearest neighbor algorithm results for the CSE-CIC-IDS2018 dataset.}
\bigskip
\begin{tabular}{ |c|c|c|c|c| } 
\hline
Data File & Accuracy & F-measure & Precision & Recall \\ 
\hline
02-14-2018.csv & 0.9999 & 0.9999 & 0.9999 & 1.0000 \\ 
\hline
02-15-2018.csv & 0.9999 & 0.9997 & 0.9996 & 0.9998 \\ 
\hline
02-16-2018.csv & 0.9999 & 0.9999 & 0.9999 & 0.9999 \\ 
\hline
02-22-2018.csv & 0.9999 & 0.9910 & 0.9822 & 1.0000 \\ 
\hline
02-23-2018.csv & 0.9998 & 0.8765 & 0.8806 & 0.8725 \\ 
\hline
03-01-2018.csv & 0.7161 & 0.4564 & 0.4960 & 0.4227 \\ 
\hline
03-02-2018.csv & 0.9999 & 0.9999 & 0.9999 & 0.9999 \\ 
\hline
All data & 0.9858 & 0.9667 & 0.9715 & 0.9621 \\ 
\hline
\end{tabular}
\end{center}
\end{table}

One can see that with the exception of the 03-01-2018.csv file the results are quite astonishing for such a simple method.  
However, the results obtained with the other more complex methods, including deep learning neural networks also show a bad performance in the case of 03-01-2018.csv \cite{key-1}-\cite{key-20}, 
with comparable values. This is due to the high similarity of attack and benign patterns for this particular data file. 
In fact, even in this extreme case the results obtained with the nearest neighbor algorithm are comparable to the 
results reported using deep learning \cite{key-20} (Convolutional Neural Networks (CNN), and Long Short-Term Memory (LSTM)), as shown in Table 3.
\begin{table}[h!]
\begin{center}
\caption{Nearest neighbor algorithm and deep learning results \cite{key-20} for the 03-01-2018.csv data.}
\bigskip
\begin{tabular}{ |c|c|c|c|c| } 
\hline
ML Method & Accuracy & F-measure & Precision & Recall \\ 
\hline
Nearest Neighbor & 0.7161 & 0.4564 & 0.4960 & 0.4227 \\ 
\hline
CNN with PCA & 0.7642 & 0.3463 & 0.7896 & 0.2218 \\ 
\hline
CNN with Autoencoder & 0.7609 & 0.3421 & 0.7595 & 0.2208 \\ 
\hline
LSTM with PCA & 0.7891 & 0.4852 & 0.7757 & 0.3530 \\ 
\hline
LSTM with Autoencoder & 0.7814 & 0.4407 & 0.7878 & 0.3059 \\ 
\hline
\end{tabular}
\end{center}
\end{table}

The CNN and LSTM results show a better accuracy and precision, however the nearest neighbor method provides a much better F-measure and recall, showing 
that it deals better with such an unbalaced data set. This result also shows that the deep learning methods overfit the data in this particular case. 

Besides the fact that the results of the nearest neighbor method are similar to those obtained using "expensive" deep learning methods, 
we should mention that the nearest neighbor approach does not actually needs training, it only needs data normalization. In contrast, the deep learning methods needed 
something like 19441.55 min (13.5 days) of training, as reported in \cite{key-2}. 

\section*{Conclusion}

Here we have investigated the machine learning approach to building an anomaly-based IDS using the CSE-CIC-IDS2018 dataset. 
We have shown that a simple nearest neighbor method provides similar results to the more "expensive" deep learning methods. 
The advantages of the nearest neighbor algorithm over the deep learning approach are: (1) it is very simple to implement; (2) it is 
extremely robust; (3) it has no parameters, and therefore it cannot overfit the data. 

In the light of the obtained results, we conclude that in the case of the CSE-CIC-IDS2018 dataset the deep learning approach (or other complex machine learning method) does not 
provide any advantage over the much more simple nearest neighbor method.

This result also shows that currently there is a trend of developing over-engineered solutions in the machine learning community. Such solutions are based on complex methods, like deep learning neural networks, 
without even considering baseline solutions corresponding to simple, but efficient methods.

\end{document}